# Controlling the strength of ferromagnetic order in $YBa_2Cu_3O_7/La_{2/3}Ca_{1/3}MnO_3$ multilayers

R. de Andrés Prada[1,2], R. Gaina[1,3], N. Biškup[4], M. Varela[4], J. Stahn[3], and C. Bernhard[1]

[1]Physics Department and Fribourg Center for Nanomaterials (FriMat), University of Fribourg, Chemin du Musée 3, CH-1700 Fribourg, Switzerland.

[2]Department of Physics, Stockholm University, AlbaNova University Center, SE-10691 Stockholm, Sweden.

[3]Laboratory for Neutron Scattering and Imaging, Paul Scherrer Institut, CH-5232 Villigen, Switzerland.

[4]Dept. de Física de Materiales & Instituto Pluridisciplinar, Universidad Complutense de Madrid, 28040 Madrid, Spain.

## I. Abstract

With dc magnetisation and polarized neutron reflectometry we studied the ferromagnetic response of $YBa_2Cu_3O_7/La_{2/3}Ca_{1/3}MnO_3$ (YBCO/LCMO) multilayers that are grown with pulsed laser deposition. We found that whereas for certain growth conditions (denoted as A-type) the ferromagnetic moment of the LCMO layer is strongly dependent on the structural details of the YBCO layer on which it is deposited, for others (B-type) the ferromagnetism of LCMO is much more robust. Both kinds of multilayers are of similar structural quality, but electron energy-loss spectroscopy (EELS) studies with a scanning transmission electron microscope reveal an enhanced average Mn oxidation state of +3.5 for the A-type as opposed to the B-type samples for which it is close to the nominal value of +3.33. The related, additional hole doping of the A-type LCMO layers, which likely originates from La and/or Mn vacancies, can explain their fragile ferromagnetic order since it places them close to the boundary of the ferromagnetic order at which even weak perturbations can induce an antiferromagnetic or glassy state. On the other hand, we show that the B-type samples allow one to obtain YBCO/LCMO heterostructures with very thick YBCO layers and, yet, strongly ferromagnetic LCMO layers.

## II. Motivation and Introduction

Multilayers based on combining superconducting and ferromagnetic materials are of considerable interest in terms of fundamental questions related to the competition between these antagonistic orders [1, 2], as well as for applications in spintronic quantum devices [3]. Recently, great progress has been made with multilayers made from conventional superconductors, like Nb or Al, and ferromagnets, like Fe, Co, or permalloy. For example, a spin-triplet SC state with Josephson-currents through micrometer-thick ferromagnetic barriers has been demonstrated [4, 5]. This raises hopes that devices based on spin-polarized supercurrents can soon be realized and integrated into a "superspintronic" technology [3]. A prerequisite for the realization of such spin-polarized superconducting/ferromagnetic quantum states is the design and precise control of the magnetic order, especially, around the interface [6].

The research on corresponding superconductor/ferromagnet multilayers from the oxide-based cuprate high $T_c$ superconductors and the perovskite manganites is comparably less advanced [7-11]. A major difficulty arises from the complexity of these oxide materials and their extremely versatile electronic and magnetic properties [12-14]. These manganites have indeed a very rich temperature and doping phase diagram with various electronic and magnetic phases that are controlled by competing interactions, like the double-exchange interaction which stabilizes a half-metallic, ferromagnetic state, and Jahn-Teller distortions which lead to insulating states with various magnetic and orbital orders [15, 16]. Even minor changes of the local structure and/or the chemical composition can lead to drastic changes of their electronic and magnetic properties. Thus, it is especially challenging to grow thin films with well-controlled properties. It is indeed well known that the ferromagnetic properties of cuprate/manganite multilayers can be strongly dependent on the choice of the substrate, the growth conditions or the layer thickness, just to name a few of the relevant factors.

Probably, the most widely studied example are multilayers from $YBa_2Cu_3O_7$ (YBCO), which in the bulk is an optimally hole-doped superconductor with $T_c$ = 90 K, and $La_{2/3}Ca_{1/3}MnO_3$ (LCMO) or $La_{2/3}Sr_{1/3}MnO_3$ (LSMO) which are half-metallic ferromagnets with bulk values of the Curie temperature of $T^{Curie}$ = 270 K and 330 K, respectively, and a saturation magnetic moment of 3.7 $\mu_B$/Mn ion [17-22]. It was found that a magnetic proximity effect occurs here at the YBCO/LCMO interface, which leads to a suppression of the ferromagnetic order on the manganite side that is commonly denoted as "dead layer" [23-25]. Concerning the strength of this suppression and the thickness of this "dead layer", the reports appear to be rather controversial. On the one hand, in superlattices made from YBCO and LCMO layers with small individual layer thicknesses of about 10 nm, it was shown that the "dead layer" can be as thin as a single or two LCMO unit cells. The suppression of the ferromagnetic moment is incomplete here and there is even a weak ferromagnetic moment induced on the YBCO side [26-29]. The ferromagnetic moment averaged of the thin LCMO layers can exceed 3 $\mu_B$/Mn ion and the Curie-temperature can be as high as 220 - 230 K [20, 30]. On the other hand, for multilayers with much thicker individual YBCO and LCMO or LSMO layers, a more pronounced "dead layer" with a complete suppression of superconductivity has been observed. Furthermore, it was reported that the thickness of the dead-layer can have a peculiar temperature dependence and increase anomalously below the superconducting transition of the YBCO layer [31-34].

The mechanism(s) responsible for this "dead layer" formation is at present only poorly understood. In addition to intrinsic effects such as interfacial strain and a transfer of electrons from LCMO to YBCO, possible extrinsic factors are chemical intermixing, the formation of oxygen or cation vacancies and other kinds of structural disorder and defects. Especially, the origin of the long-ranged proximity effect, which gives rise to a "dead layer" in excess of 10 nm, is mysterious and needs to be investigated in more detail.

In the following, we shed more light on this issue and show that the growth conditions of the LCMO layer are playing an important role in defining the thickness of this "dead layer". Moreover, we find that for certain growth conditions the structural properties of the YBCO layer can become a critical factor.

### III. Growth and Experimental

Multilayers of $YBa_2Cu_3O_7$ (Y), $Y_{0.8}Ca_{0.2}Ba_2Cu_{2.8}Co_{0.2}O_7$ (tY), $La_{2/3}Ca_{1/3}MnO_3$ (LC) and $La_{2/3}Sr_{1/3}MnO_3$ (LS) were grown by means of pulsed laser deposition (PLD) on (0 0 1)-oriented $La_{0.3}Sr_{0.7}Al_{0.65}Ta_{0.35}O_3$ (LSAT) single crystalline substrates (Crystec). Their layer sequence and thicknesses are in the following denoted according to the growth direction and in units of nanometer, i.e. as Y-15/LC-15 for a bilayer that consists of a 15 nm bottom layer of $YBa_2Cu_3O_7$ and a 15 nm top layer of $La_{2/3}Ca_{1/3}MnO_3$. For the growth, we used an excimer KrF laser ($\lambda$ = 248 nm, ts = 25 ns) with a size of the laser footprint on the target of about 3 mm$^2$. The deposition was monitored with in-situ reflection high-energy electron diffraction (RHEED) using a collimated 30 kV electron gun at low incidence angle. The substrates were glued with silver paint on a stainless steel sample holder. This holder was heated from the backside with an infrared laser and the temperature was monitored and controlled with an internal pyrometer. Before the deposition, the substrates were heated at a rate of 20 ºC·min$^{-1}$ to the deposition temperature and maintained at the growth pressure and temperature for a minimum of 60 minutes before the deposition was initiated. After deposition, the samples were cooled down at a rate of 10 ºC·min$^{-1}$. At 700ºC the oxygen pressure in the chamber was increased to 1 bar, and the cooling rate was increased to 30ºC·min$^{-1}$ until the desired in-situ annealing temperature was reached. Two different sets of growth conditions have been used in this work, which are schematically listed on Table 1.

In addition to the in-situ RHEED monitoring of the growth, the structure and composition of the samples were studied by means of x-ray diffraction (XRD), x-ray reflectivity (XRR), scanning transmission electron microscopy (STEM) and electron energy-loss spectroscopy (EELS). The XRD and XRR measurements were carried out with a commercial Rigaku Smartlab triple-axis diffractometer with a rotating-anode Cu-K$\alpha$ source ($\lambda$ = 1.54 Å). Both XRR and XRD were performed at room temperature. STEM - EELS measurements were carried out in a JEOL ARM200cF equipped with a spherical aberration corrector and a Gatan Quantum EEL spectrometer, operated at 200 kV. The specimens for STEM were prepared down the [110] orientation by conventional techniques, including mechanical grinding and Ar ion milling.

|  | **Set A of growth conditions** | **Set B of growth conditions** |
|---|---|---|
| **Deposition Temperature** | 825 °C | 840 °C |
| **Deposition Pressure** | $P_{O_2}$ = 0.30 mbar | $P_{O_2}$ = 0.34 mbar |
| **Laser fluency** | 2.00 J/cm² | 1.42 J/cm² |
| **Laser frequency** | 2 Hz | 7 Hz |
| **Post growth** | Pressure gradually increased at a constant flow of 200 sccm of $O_2$ while slowly cooling to 700 °C. | Growth pressure maintained while slowly cooling to 700 °C. |
| **In-situ annealing** | 485 °C for 1h. | 485 °C for 1h; then 400 °C for 1h. |

Table 1. Description of the two-different types of PLD growth parameters.

The magnetic properties of the samples were probed by dc magnetisation measurements using a vibrating sample magnetometer (VSM) and by polarised neutron reflectometry (PNR). The VSM data were acquired with a commercial physical properties measurement system by Quantum Design (QD-PPMS) equipped with a VSM motor and a pickup coil. The sample was fixed to a quartz rod and vibrated with a frequency of 40 Hz and peak amplitude of 2 mm. The magnetic field was applied parallel to the surface of the sample. The magnetic moment sensed by the pickup coil is a collective contribution of the sample and the substrate. Accordingly, a diamagnetic plus a weak paramagnetic signal of the LSAT substrate have been subtracted to obtain the magnetisation of the sample, $m$. The latter has been attributed to the Mn ions and, in the following, is presented in units of $\mu_B$/Mn according to the expression:

$$\boldsymbol{m}[\mu_B/_{Mn}] = \frac{\boldsymbol{m}\,[\text{emu}]}{\mu_B[\text{J}\cdot\text{T}^{-1}]} \cdot 10^{-3}\left[\frac{\text{J}}{\text{T}\cdot\text{emu}}\right] \cdot \frac{Volume\ manganite\ unit\ cell}{Volume\ manganite\ layers}$$

Polarized neutron reflectivity (PNR) measurements were performed at the AMOR beamline at the SINQ facility of the Paul Scherrer Institute (PSI) in Villigen, Switzerland. These measurements were performed in time-of-flight mode for a wavelength range of 4 Å to 13 Å using the SELENE setup [35,36]. At first, the reflectivity curves were measured in the paramagnetic state at 300K to determine the nuclear depth profile. No difference in the reflectivity curves for spin up and spin down polarization was observed here. Subsequently, the samples were measured at 90 K << $T^{Curie}$, after cooling in a magnetic field of 0.46 T applied along the in-plane direction, with polarised neutrons in spin up and spin down configurations. The obtained reflectivity curves were fitted with the program *Superfit* from the Max Planck Institute in Stuttgart. Note that for the nuclear depth profile, while the quality of the fit is fairly good, we could not find a unique solution, i.e. similarly good fits could be obtained with somewhat different fit parameters. One of the reasons seems to be that the Superfit program does not correctly account for the q-dependence of the resolution function of the SELENE setup. Note that the remaining uncertainty in the nuclear depth profile does not strongly affect the fitting of the magnetic depth profile, which is governed by the asymmetry of the spin up and spin down curves and thus is less sensitive to the resolution function and probable shortcomings of the structural model.

IV.     Results and Discussion

**IV.A. X-ray Diffraction (XRD) and Reflectometry (XRR) Study**

Figure 1 compares the XRD and XRR curves of two LC-15/Y-15 and Y-15/LC-15 bilayers that were both deposited with the A-type growth conditions (see Table 1). The only difference between these two bilayers concerns the stacking sequence of the YBCO and LCMO layers. The Θ-2Θ scans in Figs. 1(a) and 1(b) reveal (0 0 *l*) Bragg peaks that are reasonably strong and narrow, and confirm the epitaxial and c-axis oriented growth of the layers. No traces of misoriented material or of impurity phases can be found. Figure 1(c) details the region around the LSAT (0 0 1), LCMO (0 0 1) and YBCO (0 0 3) Bragg peaks which reveal clear satellites that are characteristic of a high crystalline quality and coherence of the layers. Finally, Figs. 1(d) and 1(e) show the XRR curves obtained on each bilayer. Pronounced Kiessig fringes occur for both bilayers, indicating that the interfaces between the different layers are reasonably sharp and coherent. From the fits (red lines) we obtained the thickness and roughness parameters of the layers as listed in the insets.

Next, Figure 2 compares the XRD curves of two Y-20/LC-30 bilayers that were grown with the different PLD conditions A and B. Figures 2(a) and 2(b) compare the $\Theta$-$2\Theta$ scans for the A-type (green) and B-type (red) bilayers, respectively. For both growth types, they confirm a uniform orientation of the films along the $c$ crystalline direction, without traces of impurity phases or parts that are misoriented. The close-up of the LCMO (0 0 1) and YBCO (0 0 3) peaks in Figure 2(c) reveals that for both the A-type and B-type bilayers there exist similarly pronounced satellite peaks that testify for a high crystalline quality of the YBCO and LCMO layers. The main difference concerns a very small shift of the centre of the peak which occurs at a slightly smaller angle for the A-type sample.

**IV.B. Scanning transmission electron microscopy (STEM)**

One B-type and three A-type YBCO/LCMO multilayers have been investigated by means of atomic resolution STEM-EELS techniques. Figures 3 and 4 show representative results for the B-type Y-22/LC-30 and an A-type Y-15/LC-20 bilayer.

High resolution STEM high angle annular dark field (HAADF) images are compared in Figures 3(a) and 3(b), showing the YBCO layers for the B-type and A-type case, respectively. A black arrow highlights the position of the LCMO/YBCO interface for both images. The blemished spot in the middle of Fig. 3(a) is an artifact consequence of sample preparation (ion milling) and is not inherent to the as-grown sample. The interfaces appear coherent and mostly defect free. Both A- and B-type YBCO layers exhibit a number of stacking faults in the form of Y-124-like double or even triple CuO chains (darker horizontal atomic planes) [37, 38]. In the case of the A-type sample in Fig. 3(b) these stacking faults are quite evenly distributed throughout the 15 nm thick YBCO layer. For the thicker B-type YBCO in Fig. 3(a) they are concentrated within the first 14 nm above the substrate whereas the last 7 atomic layers of YBCO, i.e. those close to the LCMO interface, exhibit the expected Y-123 structure. This finding indicates that the formation of these stacking faults and the related non stoichiometric growth might be enhanced by the strain relaxation of the YBCO layer, which has been reported to occur within the first 15 nm [30, 39].

Figures 3(a) - 3(d) reveal that the LCMO layers of both the B-type and A-type samples exhibit a good degree of crystallinity. Nevertheless, the analysis of EELS linescans including the Mn $L_{2,3}$ absorption edge, which have been acquired along the out-of-plane direction as indicated by the thick yellow bars in Figs. 3(c) and 3(d), suggests clear differences in the electronic properties of the LCMO layers. Note that the linescan of the B-type sample in Fig. 3(c) does not span the entire 30 nm thick LCMO layer but tracks the manganese signal (in our case the Mn $L_{2,3}$ white line at an approximate energy loss of 640 eV) only in a range of 10 nm from the YBCO/LCMO interface. The manganese $L_{2,3}$ intensity ratio depends on the oxidation state of manganese ion, and can be used for a quantification [38]. The Mn $L_{2,3}$ ratio of the two samples, derived with a double derivative method [40], is plotted in Fig. 3(e). Black squares and red circles denote the B-type and A-type samples, respectively. The right axis shows the corresponding Mn oxidation state based on the calibration of bulk manganite samples. The oxidation state of manganese on the B-type sample across the B-type manganite layer is close to the nominal value of +3.33 as expected for this compound, while in the A-type sample it has a significantly higher value of about +3.5. Correspondingly, high values for the manganite oxidation state have been observed for the other two A-type samples. A manganese oxidation state of +3.5 suggests the possibility of a non stoichiometric chemical composition of the A-type $La_{1-x}Ca_xMnO_3$ layers, which would lead to an enhanced hole doping to a level comparable to the one of $La_{0.5}Ca_{0.5}MnO_3$, i.e. close to the antiferromagnetic border in the $La_{1-x}Ca_xMnO_3$ phase diagram [15, 16]. Since an excess oxygen concentration is typically not observed in these manganese perovskites, the most likely cause could be the presence of Mn and/or La vacancies [41-45]. Especially the Mn vacancies also give rise to disorder effects that can suppress the ferromagnetic order.

Figure 4 shows a number of atomic resolution EELS images of the LCMO/YBCO interface of the A-type bilayer. The right panel exhibits a high magnification HAADF image of the interface. A yellow rectangle depicts the region where an EELS spectrum image was taken. The simultaneously acquired ADF signal and elemental maps based on the analysis of the O K, Mn $L_{2,3}$, Ba $M_{4,5}$, La $M_{4,5}$ and Cu $L_{2,3}$ edges are shown on the left panels (respectively, from left to right). In spite of some spatial drift, the atomic columns are clearly resolved. The elemental profiles, obtained by horizontal averaging of each two dimensional elemental map are shown on the survey ADF image. The atomic plane stacking of the LCMO/YBCO interface can be extracted from these data; it is such that a manganite $MnO_2$ plane from the LCMO faces a BaO plane from YBCO, as is commonly observed in this system [28, 30, 46, 47]. Yellow, green and orange points in the ADF images depict the positions of Ba, La and Mn atoms at the LCMO/YBCO interface. The interface itself is indicated by a yellow dashed line. This type of interface termination has been found in both the A- and B-type samples, although in the former sample a different, $LaO-CuO_2$ termination has also been detected on occasion. This second type of termination has not been found for the B-type sample, but we cannot discard its existence due to the lack of statistics of the technique.

Accordingly, the most important difference between the A- and B-type bilayers concerns the average manganese oxidation state which is close to the nominal one of +3.33 for the latter but clearly enhanced to about +3.5 for the former case. On the other hand, for both types of bilayers, the LCMO layers are of a comparable structural quality and the YBCO/LCMO interfaces are well defined and atomically sharp with no sign of chemical intermixing. For both bilayers the typical $MnO_2$-BaO-$CuO_2$ interlayer stacking sequence dominates, albeit for the A-type bilayer a different, $MnO_2$-LaO-$CuO_2$ termination has been detected occasionally. The YBCO layers of the A- and B-type samples contain stacking faults in terms of CuO bilayers that seem to be more abundant in the strain relaxation region within about 15 nm of the LSAT substrate. The limited statistics of the STEM data does not allow us to draw firm conclusions about the differences between the A- and B-type samples concerning these YBCO stacking faults and the minority interfacial layer stacking sequence.

**IV.C. Magnetic Properties of LCMO Layers**

**IV.C.1. DC Magnetisation of A-type Y-15/LC-15 and LC-15/Y-15 with inverted stacking sequence**

Figure 5 shows the magnetisation data for the 15/15 bilayers described in Section IV.A. The magnetisation curves reveal that the ferromagnetic signal is much stronger for the LC-15/Y-15 bilayer than for the Y-15/LC-15 one. The dc magnetisation data in Fig. 5(a), which were taken in field-cooling mode with an external field of $H_{ext}$ = 1 kOe applied parallel to the layers, show for LC-15/Y-15 a pronounced ferromagnetic transition around $T^{Curie} \approx 200$ K with a spontaneous magnetisation at 90 K that reaches about 2.75 $\mu_B$/Mn ion. Note that the magnetisation data below 90 K are strongly affected by the superconducting response of the YBCO layer and are therefore not further discussed here. In stark contrast, the magnetic signal of the Y-15/LC-15 bilayer shows no clear sign of a ferromagnetic transition and increases only gradually toward low temperature, reaching a value of only 0.12 $\mu_B$/Mn ion at 90 K that is more than twenty times smaller than the one of LC-15/Y-15. This marked suppression of the ferromagnetic moment of the Y-15/LC-15 bilayer is also evident from the magnetisation-field (*M-H*) loops at 100 K shown in Fig. 5(b). The LC-15/Y-15 bilayer shows a pronounced ferromagnetic signal with a high field saturation value of about 2.8 $\mu_B$/Mn ion and a pronounced hysteresis at low field with a coercive field of $H_{coer} \approx 70$ Oe and a remnant magnetisation of $M_{rem} \approx 1.8$ $\mu_B$/Mn ion. The signal of the bilayer Y-15/LC-15 is again extremely weak with no clear sign of a hysteresis and a saturation value of only 0.2 $\mu_B$/Mn ion. The data in Figure 5 highlight that the ferromagnetic properties of these 15 nm-thick A-type LCMO layers depend dramatically on whether they are grown directly on the LSAT substrate or on top of an intermediate 15 nm-thick YBCO layer. We have also verified that the magnetic properties of an A-type LSAT/LC-15 single layer without YBCO (data not shown) are very similar to the ones of the LC-15/Y-15 bilayer and also to the ones reported in the literature for similarly thin LCMO layers that are directly grown on substrates like STO, NGO or even SLAO [30, 48].

### IV.C.2. DC Magnetisation of A-type structures with different thickness of intermediate YBCO

The decisive role of the intermediate YBCO layer is further highlighted in Fig. 6, which compares the magnetic response of a Y-15/LC-15/Y-30 trilayer with the one of a Y-10/LC-15/Y-30 trilayer. Both samples are almost identical, except for the different thickness of the first YBCO layer of $d^Y \approx 15$ nm and 10 nm, respectively. Figure 6 reveals that this seemingly small difference in the thickness of the intermediate YBCO layer has quite a dramatic effect on the ferromagnetic response of the A-type LCMO layer that is grown on top. For the Y-

10/LC-15/Y-30 trilayer (blue symbols) the 1 kOe field-cooled curve in Fig. 6(a) shows a clear ferromagnetic transition with $T^{Curie} \approx 180$ K and a sizeable magnetic moment of 1.65 $\mu_B$/Mn ion at 90 K. Likewise, the *M-H* loop in Fig. 6(b) shows a typical ferromagnetic response with a saturation value of 1.7 $\mu_B$/Mn at 5 Tesla and a clear hysteresis around the origin with a coercive field of $H_{coer} \approx 40$ Oe and a remnant magnetisation of $M_{rem} \approx 0.9$ $\mu_B$/Mn ion. To the contrary, the Y-15/LC-15/Y-30 trilayer (pink symbols) shows a very weak ferromagnetic response, similar to the Y-15/LC-15 bilayer in Fig. 5, with no sign of a hysteresis around the origin and an extremely small saturation moment of 0.12 $\mu_B$/Mn ion.

This finding raises the question about the relevant changes of the properties of the YBCO layer that occur upon such a seemingly moderate reduction of the thickness of the YBCO layer from $d^Y$ = 15 nm to 10 nm. An important factor appears to be the strain relaxation of the YBCO layer that has been reported to occur around a critical thickness of $d_{crit}^Y \approx 15$ nm [30, 39]. This strain relaxation typically gives rise to the formation of various structural defects, like stacking faults or screw dislocations. Moreover, it can affect the orthorhombic distortion of the fully oxygenated YBCO, due to the one-dimensional CuO chains along the b-axis direction, which most likely remains very small (or short ranged) for the fully strained (unrelaxed) part of the YBCO layer for which the in-plane structure is locked to the one of the cubic LSAT substrate. The length and ordering of the CuO chain segments and the magnitude of the related orthorhombic distortions is likely to increase rapidly as the strain relaxation sets in and the YBCO layer develops its intrinsic structure.

**IV.C.3. DC Magnetisation of A- and B-type Y-20/LC-30 bilayers**

Figure 7 shows that a modification of the growth condition of these multilayers can also give rise to surprisingly large changes of the ferromagnetic properties of the LCMO layers grown on top of YBCO. It displays the magnetisation data of two Y-20/LC-30 bilayers that were PLD-grown with the conditions A and B as described in Section III. Whereas the superconducting transition temperature of $T_c \approx 80$ K of the YBCO layer is very similar (data not shown), their ferromagnetic properties are dramatically different. For the A-type bilayer, the magnetisation curve in Fig. 7(a) increases only very gradually toward low temperature and reaches a value of only about 0.3 $\mu_B$/Mn ion at 90 K. The corresponding *M-H* loop at 100 K in

Fig. 7(b) also shows no sign of a hysteresis around the origin and saturates at 0.3 $\mu_B$/Mn ion. To the contrary, the B-type bilayer exhibits a strong ferromagnetic response with a pronounced transition at $T^{Curie} \approx 210$ K and a magnetic moment of 2.8 $\mu_B$/Mn at 90 K in Fig. 7(a). The M-H loop at 100 K in Fig. 7(b) confirms a large saturation value of 2.8 $\mu_B$/Mn and exhibits a pronounced hysteresis around the origin with a coercive field of $H_{coer} \approx 140$ Oe and a remnant magnetisation of $M_{rem} \approx 2.15$ $\mu_B$/Mn ion.

**IV.C.4 Polarized Neutron Reflectometry of A- and B-type Y-20/LC-30 bilayers**

The magnetic depth profile of the same Y-20/LC-30 bilayers, for which the dc magnetisation is shown in Fig. 7, has been investigated with polarized neutron reflectometry (PNR). Figures 8(a) and 8(b) show the reflectivity curves at 300 K (symbols) and the best fit (solid line) from which the depth profile of the nuclear scattering length density, the thickness and the roughness of the LCMO and YBCO layers have been determined as listed in Table 2. Figures 8(c) and 8(d) show the corresponding curves and fits for spin-up and spin-down polarized neutrons in the ferromagnetic state at 90 K. For the B-type bilayer in Fig. 8(d) there is a pronounced splitting between the reflectivity curves for the spin-up and spin-down neutrons that is characteristic of a rather large magnetisation of the LCMO layer. For the A-type bilayer in Fig. 8(c) this splitting is very weak and barely visible. To obtain the depth profile of the magnetic scattering length density of these bilayers from the spin up and spin down curves at 90 K, the parameters describing the nuclear part were fixed to the ones obtained from the 300 K curves. Figures 8(e) and 8(f) show the magnetic depth profiles (blue lines) that have been obtained from the best fits that are displayed by the solid lines in Figs. 8(c) and 8(d). For the A-type bilayer they confirm a strong suppression of the ferromagnetic moment of the LCMO layer and yield a dead layer with a thickness on the order of about 20 nm and an average moment of only 0.30 $\mu_B$/Mn ion. To the contrary, for the B-type bilayer they yield a large ferromagnetic moment with an average value of 3.27 $\mu_B$/Mn that agrees well with the one deduced from dc magnetisation (see Fig. 7). Furthermore, this B-type bilayer has only a thin "dead layer" at the YBCO/LCMO interface of about 5.5 nm for which the magnetisation is reduced but not even fully suppressed.

|   | **A-type** | | | **B-type** | | |
| --- | --- | --- | --- | --- | --- | --- |
| Layer | Thickness (nm) | SLD ($10^{14}$ m$^{-2}$) | m ($\mu_B$/Mn) | Thickness (nm) | SLD ($10^{14}$ m$^{-2}$) | m ($\mu_B$/Mn) |
| LCMO$_6$ | 2.70 ± 0.25 | 3.65 | 0.10 | 1.90 ± 0.30 | 3.48 | 2.47 |
| LCMO$_5$ | 8.00 ± 0.20 | 3.65 | 0.80 | 3.70 ± 0.40 | 3.60 | 2.74 |
| LCMO$_4$ | 5.60 ± 0.26 | 3.50 | 0.64 | 8.00 ± 0.33 | 3.60 | 3.65 |
| LCMO$_3$ | 7.50 ± 0.13 | 3.50 | 0.10 | 8.00 ± 0.32 | 3.60 | 3.60 |
| LCMO$_2$ | 5.50 ± 0.30 | 3.65 | 0.00 | 7.50 ± 0.10 | 3.37 | 3.55 |
| LCMO$_1$ | 7.00 ± 0.06 | 3.65 | 0.00 | 5.50 ± 0.20 | 3.37 | 2.50 |
| YBCO$_3$ | 11.00 ± 0.20 | 4.20 | | 6.30 ± 0.25 | 4.47 | |
| YBCO$_2$ | 5.00 ± 0.10 | 4.65 | | 4.10 ± 0.10 | 4.74 | |
| YBCO$_1$ | 5.00 ± 0.10 | 4.65 | | 10.00 ± 0.24 | 4.54 | |
| **LSAT substrate** | | | | | | |

**Table 2.** Parameters used to fit the polarized neutron reflectometry data of the A-type and B-type Y-20/LC-30 bilayers.

**IV.C.5 Enhanced DC Magnetisation for multilayers with cation-substituted, tetragonal YBCO**

To test the hypothesis that the orthorhombic distortions due to the long-range order of the CuO chains of YBCO contribute to the strong suppression of the ferromagnetic order of LCMO layers that are grown on top, we studied corresponding multilayers with intermediate $Y_{0.8}Ca_{0.2}Ba_2Cu_{2.8}Co_{0.2}O_7$ layers. The Co$^{3+}$ ions are incorporated mainly within the Cu-O chains and give rise to a strong structural disorder that suppresses the orthorhombicity of YBCO. For bulk YBCO crystals, a quasi-tetragonal structure without any long range order of the CuO chains is thus obtained for a Co content $y > 0.15$ [49]. The additional partial substitution of Ca$^{2+}$ on the Y$^{3+}$ site is used to counteract the electron doping due to the Co$^{3+}$ substitution, and thus to maintain an almost optimal doping level with a reasonably high superconducting transition temperature of $T_c \approx 73$ K for these tetragonal $Y_{0.8}Ca_{0.2}Ba_2Cu_{2.8}Co_{0.2}O_7$ (tYBCO) layers.

Figure 9 compares the magnetisation data of the A-type Y-15/LC-15 bilayer, which are also shown in Fig. 5 and discussed in Section IV.C.1, with the ones of an A-type grown tY-30/LC-15 bilayer. It is evident that the latter tY-30/LC-15 bilayer exhibits a much stronger ferromagnetic signal than the former Y-15/LC-15 bilayer for which the ferromagnetic signal is extremely weak. The ferromagnetic signal arising in tY-30/LC-15 in Figure 9(a) demonstrates a magnetic activation at temperature of about $T^{Curie} \approx 180$ K that reaches 0.45 $\mu_B$/Mn at 90 K.

Furthermore, a hysteresis is found around zero field in the in-plane *M-H* curves obtained at 100 K, Figure 9(b), with a finite coercive field of $H_{coer} \approx$ 55 Oe and a remnant magnetisation of $M_{rem} \approx 0.14$ $\mu_B$/Mn ion. The moment rapidly saturates with the field (see inset in Fig. 9(b)), reaching 0.38 $\mu_B$/Mn ion at 1.5 T. Note that this enhancement of the ferromagnetic response occurs despite of the much larger (doubled) thickness of the intermediate tetragonal YBCO layer which for the case of orthorhombic YBCO has been shown to be detrimental to the magnetic properties of a LCMO layer that is grown on top (see Fig. 6 and the discussion in Section IV.C.2). This comparison thus demonstrates that the suppression of the ferromagnetic signal of the LCMO layer is less pronounced if it is grown on top of a tetragonal YBCO (tY) layer instead of regular orthorhombic YBCO.

Figure 10 shows that the tetragonal YBCO can also be used in B-type multilayers to enhance the ferromagnetism of thin LCMO layers that are grown on top of a very thick YBCO layers. It compares the magnetisation curves of a Y-100/LC-10/Y-100 trilayer (red symbols) and a corresponding tY-100/LC-10/Y-100 trilayer (yellow symbols) that are both B-type grown. Figure 10(a) shows that, whereas both trilayers have similar Curie temperatures of $T^{Curie} \approx$ 200 K, the magnetisation of tY-100/LC-10/Y-100 is considerably larger with 1.75 $\mu_B$/Mn at 90 K than for Y-100/LC-10/Y-100 with 0.9 $\mu_B$/Mn at 90 K. A corresponding difference is also evident for the *M-H* curves in Fig. 10(b), for which both trilayers exhibit a pronounced hysteresis around the origin with coercive fields of $H_{coer} \approx$ 360 Oe and 220 Oe and remnant magnetisations of $M_{rem} \approx 0.7$ $\mu_B$/Mn ion and 1.3 $\mu_B$/Mn ion for trilayers with the intermittent Y-100 and tY-100 layers, respectively. The rather strong enhancement of the coercive field of Y-100/LC-10/Y-100 is typical for multilayers for which a thin LCMO layer is grown on top of a very thick YBCO layer and is likely also related to the orthorhombic distortion of the latter. The inset of Fig. 10(b) shows that the high field saturation value amounts to 0.95 $\mu_B$/Mn for Y-100/LC-10/Y-100 as compared to 1.75 $\mu_B$/Mn for tY-100/LC-10/Y-100.

**IV.C.6 Superconducting spin-valve structures with very thin manganite layers and large, switchable magnetisation**

Last but not least, we show in Fig. 11 the corresponding dc magnetisation data of a more complex multilayer structure with very thick superconducting layers and two very thin LCMO and LSMO ferromagnetic layers. The profile of the multilayer is tY-100/LS-5/Y-4/LC-10/Y-100 according to the formalism introduced at the beginning of Section III. This sample was made in an attempt to grow a superconducting spin-valve structure for which two thick superconducting (YBCO and tYBCO) electrode-layers are separated by two very thin ferromagnetic layers that have a strong magnetic response with different coercive fields, such that their magnetisation direction can be individually switched with an external magnetic field. Figure 11 confirms that we managed to obtain a strong ferromagnetic order within the LCMO and LSMO layers, even though they are very thin and grown on top of a 100 nm-thick tYBCO layer. Figure 11 (a) shows that the *M-T* curve of this multilayer has two ferromagnetic components, one that develops already above 300 K due to the LSMO layer with $T^{Curie}$ > 300 K, and an additional one that appears below $T^{Curie} \approx 210$ K and can be assigned to the LCMO layer. The *M-H* loop in Fig. 11(b) confirms that the LCMO and LSMO layers have different coercive fields of about 300 Oe and 40 Oe, respectively. Their magnetisation can indeed be independently switched, as is seen by the two step behaviour of the hysteresis loop that is most clearly seen in the derivative plot of *dM/dH*. This is a promising first step toward the growth of SC/FM multilayers from these complex oxides that can be used to study the spin-valve effect and related quantum spintronic phenomena.

## V.    Discussion and Summary

The results described above reveal that the ferromagnetic properties of a LCMO layer that is grown with pulsed laser deposition (PLD) on top of the cuprate high $T_c$ superconductor $YBa_2Cu_3O_7$ (YBCO) are very sensitive to the growth conditions and strongly dependent on the structural details of the YBCO layer.

We have shown that thin LCMO layers that are prepared with the A-type conditions (described in Table 1) exhibit a strong ferromagnetic response if grown directly on a LSAT substrate, whereas their ferromagnetic moment is drastically reduced when they are grown on an intermediate YBCO layer. Moreover, Fig. 6 reveals that the thickness of the intermediate YBCO layer plays an important role in the suppression of the ferromagnetic moment of the A-type LCMO layers. This suggests that the dislocations and defects related to the strain relaxation that sets in near a critical YBCO layer thickness of about 15 nm [30, 39] are detrimental to the ferromagnetic order of the LCMO layer on top. Likewise, we have obtained evidence for a negative impact due to the orthorhombic structure of YBCO that arises when the one-dimensional CuO chains develop a long-range order. This hypothesis has been confirmed by replacing the orthorhombic YBCO with tetragonal $Y_{0.8}Ca_{0.2}Ba_2Cu_{2.8}Co_{0.2}O_7$ (tY), for which the CuO chains are strongly disrupted by the Co ions (which prefer a three-fold oxygen coordination). The data in Fig. 9 confirm that the ferromagnetic moment of an A-type tY-30/LC-15 bilayer is more than doubled as compared to an A-type Y-15/LC-15 bilayer with only half the YBCO layer thickness.

Moreover, we have shown that by using the modified B-type growth conditions (detailed in Table 1) one can obtain LCMO layers with a strong ferromagnetic order that is less sensitive to the structural details of the intermediate YBCO layer. The dramatic differences between the ferromagnetic properties of such A- and B-type multilayers have been demonstrated for a pair of Y-20/LC-30 bilayers, for which the average magnetisation and the magnetic depth profile have been measured with dc magnetisation and polarized neutron reflectometry (PNR). For the B-type bilayer they revealed a bulk-like ferromagnetic order with an average magnetic moment of 3.27 $\mu_B$/Mn ion at 90 K without a fully suppressed ferromagnetic moment at the LCMO/YBCO interface. In stark contrast, for the A-type bilayer they revealed only a very weak ferromagnetic response with an average moment of 0.30 $\mu_B$/Mn ions and a thick dead layer of about 20 nm next to the YBCO/LCMO interface for which the ferromagnetic order is entirely absent.

Finally, we have demonstrated that using a B-type growth mode and tetragonal $Y_{0.8}Ca_{0.2}Ba_2Cu_{2.8}Co_{0.2}O_7$ (tYBCO) layers allows one to obtain very thin LCMO layers (10 nm) with a strong ferromagnetic response on top of thick tYBCO layers (100 nm). Last but not least, the data in Fig. 11 highlight that this approach enables the growth of spin-valve structures with thick YBCO electrodes and, yet, two very thin manganite layers with sizeable ferromagnetic moments whose relative orientation can be readily adjusted with an external magnetic field.

These observations raise the important question as to why the ferromagnetic properties of the A-type LCMO layers are so strongly affected by the structural details of the intermediate YBCO layer on which they are grown, whereas the ferromagnetism of the B-type samples is much more robust. The structural analysis based on the XRD and TEM data suggests that both the A- and B-type multilayers have a high structural quality with epitaxial YBCO and LCMO layers that have flat and sharp interfaces and show no signs of chemical inter-diffusion. The most pronounced difference between these A- and B-type samples is identified from the analysis of the STEM-EELS data and concerns the average oxidation state of the Mn ions. Whereas for the B-type samples it is close to the nominal value of +3.3, for the A-type samples it is strongly enhanced up to about +3.5. Such a pronounced increase of the Mn oxidation state and thus of the hole doping of the A-type LCMO layers arises most likely from a non-ideal cation stoichiometry due to a deficiency of the La and/or Mn ions. A corresponding cation deficiency has also been observed in nominally undoped $LaMnO_3$ thin film samples for which the additional hole doping gives rise to a ferromagnetic order instead of the antiferromagnetic state of the undoped parent compound [41]. For the present A-type LCMO layers the cation deficiency weakens the ferromagnetic order in two ways. Firstly, it enhances the hole doping and thus places them in the doping phase diagram close to a phase boundary that separates the itinerant ferromagnetic order at $x < 0.5$ from antiferromagnetic and charge/orbital ordered states at $x > 0.5$ [15, 16]. The ferromagnetic properties near this phase boundary are known to be very sensitive to defects and lattice distortions which reduce the band width and thus the strength of the double-exchange interaction. Secondly, the Mn-vacancies are directly disrupting the network of Mn-O-Mn bonds and thereby further reduce the effective band-width and the ferromagnetic double-exchange interaction, thus favouring antiferromagnetic and/or glassy types of magnetic orders [41, 15, 16].

In summary, we have shown that the ferromagnetic order of LCMO layers that are grown with pulsed laser deposition (PLD) on an intermediate YBCO layer can be strongly dependent on the PLD growth conditions. A moderate variation of the growth parameters can result in a significant cation deficiency, which increases the hole doping of the LCMO layer and thus places it close to a phase boundary in the doping phase diagram at which the ferromagnetic order is easily suppressed in favour of an antiferromagnetic and/or glassy state. Even minor strain effects and defects that arise from the relaxation and the orthorhombic distortion of the underlying YBCO layer can therefore give rise to a strong suppression of the ferromagnetic order and induce a very thick dead layer. Furthermore, we have demonstrated that with PLD one can grow samples with stoichiometric LCMO layers with the nominal hole doping close to $x$ = 0.33, for which the ferromagnetic order is very robust and less dependent on the structural details of the underlying YBCO layer. Very thin (10 nm) and strongly ferromagnetic LCMO layers thus can be readily grown on top of thick (100 nm) YBCO layers. Even superconducting spin-valve structures with two such thin ferromagnetic manganite layers can be prepared, for which the magnetisation direction can be switched independently with an external magnetic field. Such superconductor/ferromagnet heterostructures with very thick bottom electrodes and, yet, very thin ferromagnetic layers with sizeable and individually switchable magnetic moments are of great interest in fundamental and applied science. For certain non-collinear magnetic arrangements they allow one to induce a superconducting order parameter with a large spin-triplet component that can be a source of spin polarized supercurrents for superspintronic devices or memory cells [1-6].

**Acknowledgements:** This work is partially based upon experiments performed at the Swiss spallation neutron source SINQ, Paul Scherrer Institute, Villigen, Switzerland. The work at the University of Fribourg was supported by the Swiss National Science Foundation (SNF) through grants No. 200020-172611 and CRSII2-154410/1. The electron microscopy studies were carried out at the Centro Nacional de Microscopía Electrónica at the University Complutense de Madrid and sponsored by the Spanish MINECO/MICINN/FEDER grants MAT2015-66888-C3-3-R and RTI2018-097895-B-C43.

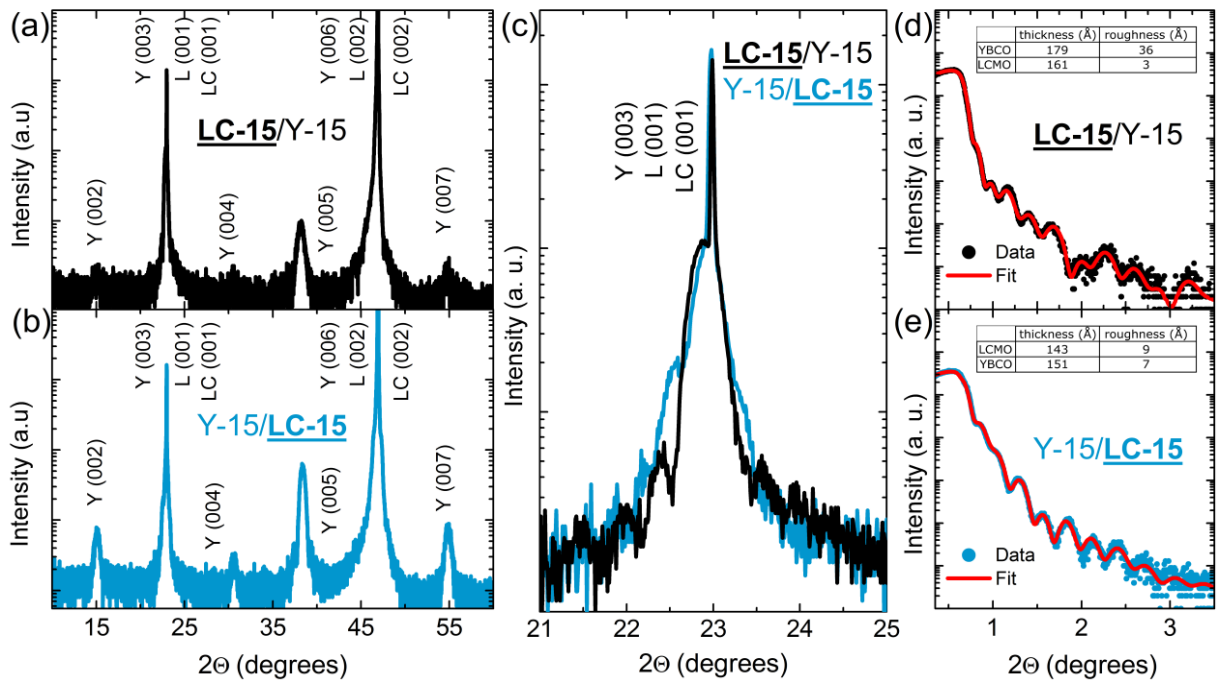

**Figure 1.** X-ray data of LC-15/Y-15 (black lines) and Y-15/LC-15 (blue lines) bilayers grown with the A-type conditions described in Table 1. **(a)** and **(b)** The Θ-2Θ scans of the LC-15/Y-15 and Y-15/LC-15 bilayers, respectively. **(c)** Detailed view of the overlapping LSAT (0 0 1), LCMO (0 0 1), and YBCO (0 0 3) Bragg-peaks. The presence of satellite peaks demonstrates the long-range coherence of the crystalline planes. **(d)** and **(e)** X-ray reflectivity (XRR) curves (symbols) with the respective fits (red lines) and the deduced thickness and roughness parameters. The Bragg-peaks corresponding to each material are labelled as *Y* for YBCO, *L* for LSAT and *LC* for LCMO.

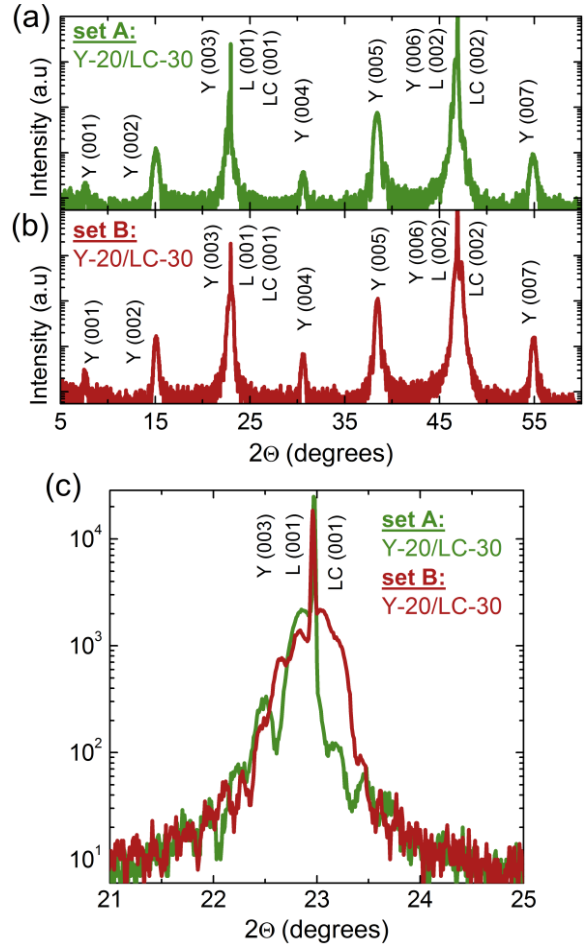

**Figure 2.** X-ray data of Y-20/LC-30 bilayers grown with the A-type (green lines) and B-type (red lines) conditions. **(a)** and **(b)** The Θ-2Θ scans of the type-A and type-B bilayers, respectively. **(c)** Detailed view of the overlapping LSAT (0 0 1), LCMO (0 0 1), and YBCO (0 0 3) Bragg-peaks. Satellite peaks are observed for both bilayers, which demonstrate a long-range coherence of the crystalline planes. The Bragg-peaks corresponding to each material are labelled as *Y* for YBCO, *L* for LSAT and *LC* for LCMO.

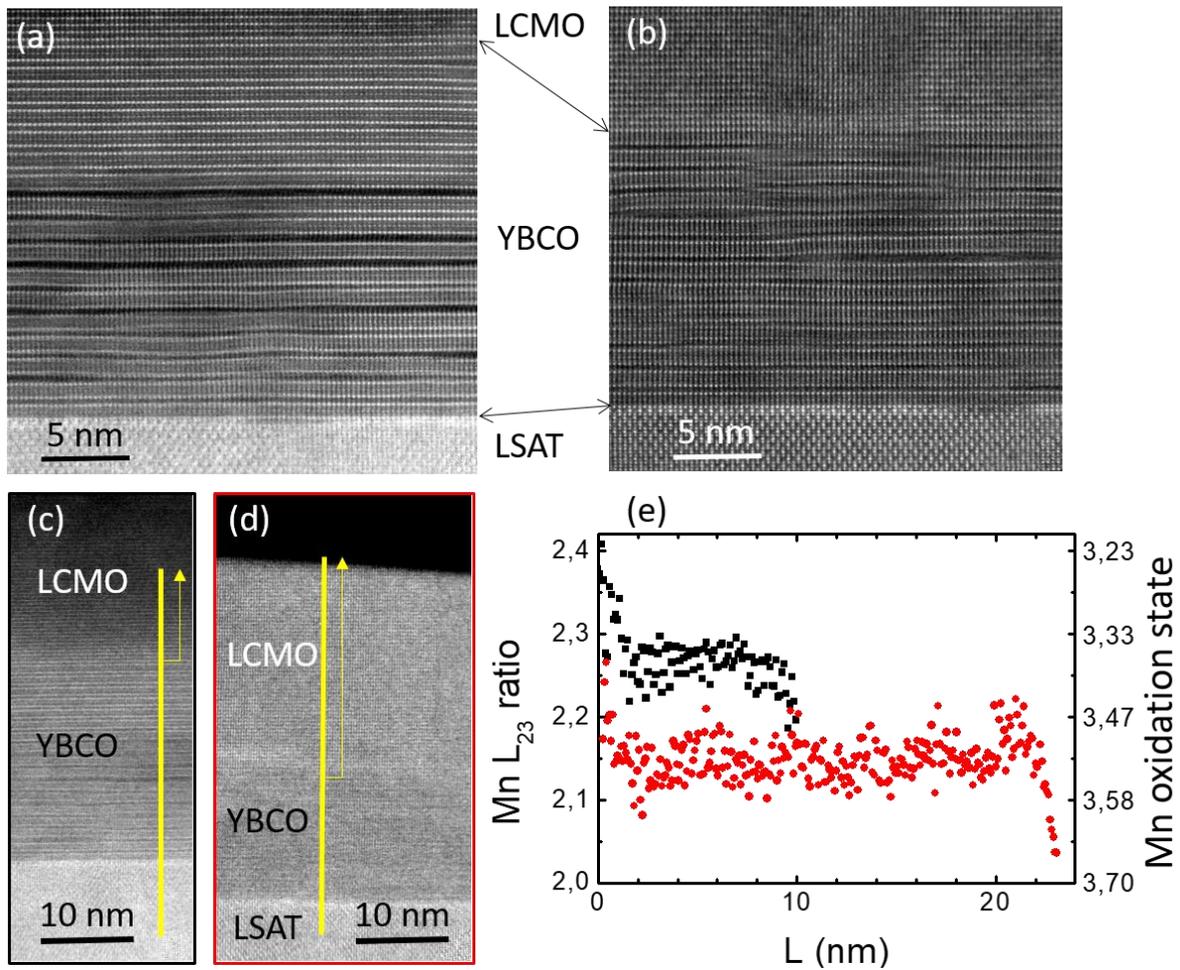

**Figure 3.** High resolution ADF images of **(a)** a B-type Y-22/LC-30 and **(b)** an A-type Y-15/LC-20 bilayer, respectively. The LSAT-YBCO and YBCO-LCMO interfaces are marked by arrows. Panels **(c)** and **(d)** show the survey images for the EELS linescans across the layers. Panel **(e)** shows the manganese $L_{2,3}$ ratio across the LCMO layer for the B-type (black squares) and the A-type bilayers (red circles). The position where $L$ = 0 nm has been chosen to indicate the location of the LCMO/YBCO interface and the distance $L$ is measured in the direction of the arrows in panels **(c)** and **(d)**.

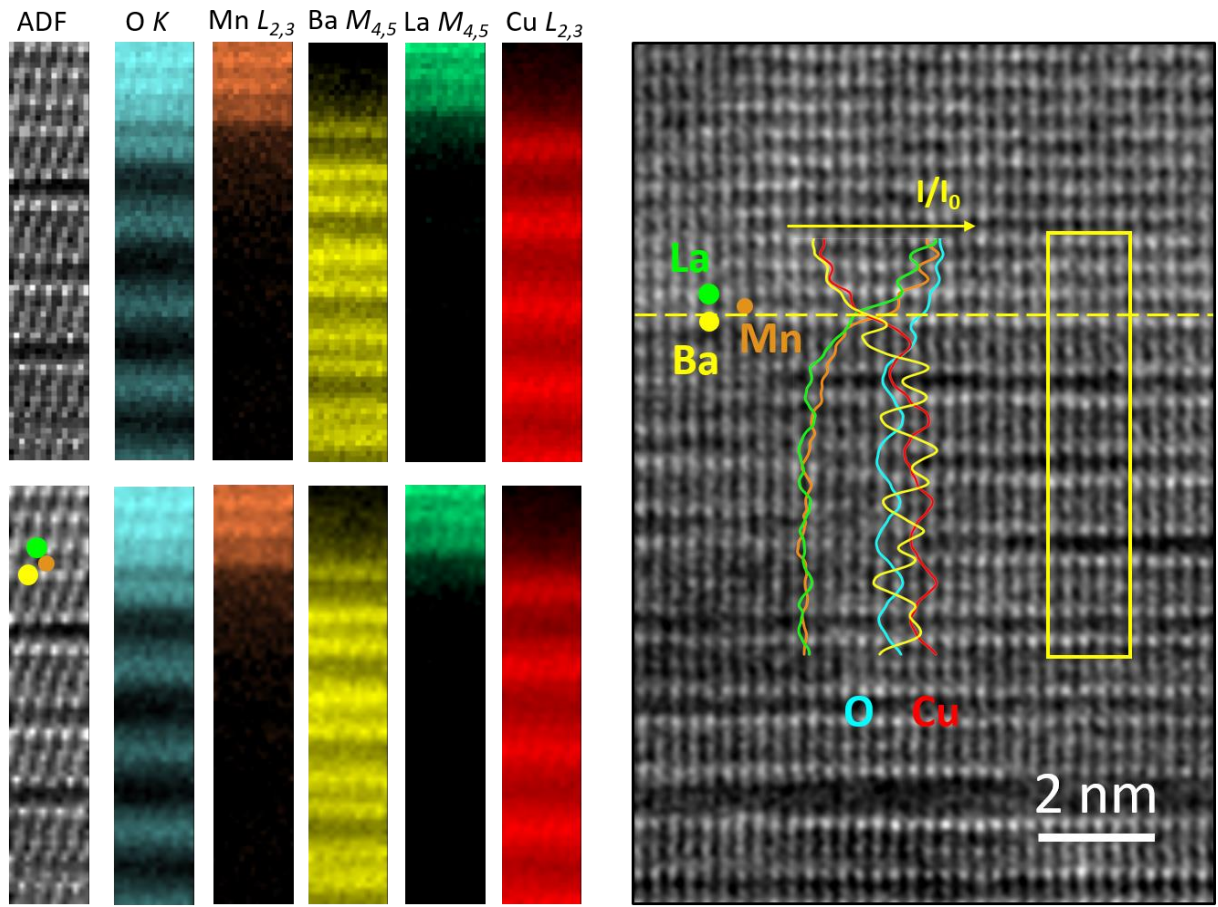

**Figure 4. Right panel:** the EELS survey image on the A-type bilayer showing the region where EELS data are acquired (yellow rectangle). The LCMO/YBCO interface is marked by the dashed yellow line, and three particular atoms of La, Mn and Ba at the interface are indicated by circles. **Left panel:** the simultaneous ADF image and elemental maps based on the O $K$, Mn $L_{2,3}$, Ba $M_{4,5}$, La $M_{4,5}$ and Cu $L_{2,3}$ edges (**top:** raw images, **bottom:** smoothed data). The inset in the right panel shows the laterally averaged profiles corresponding to these two-dimensional maps, in the same color scale. Random noise in the EELS data was removed by means of principal component analysis.

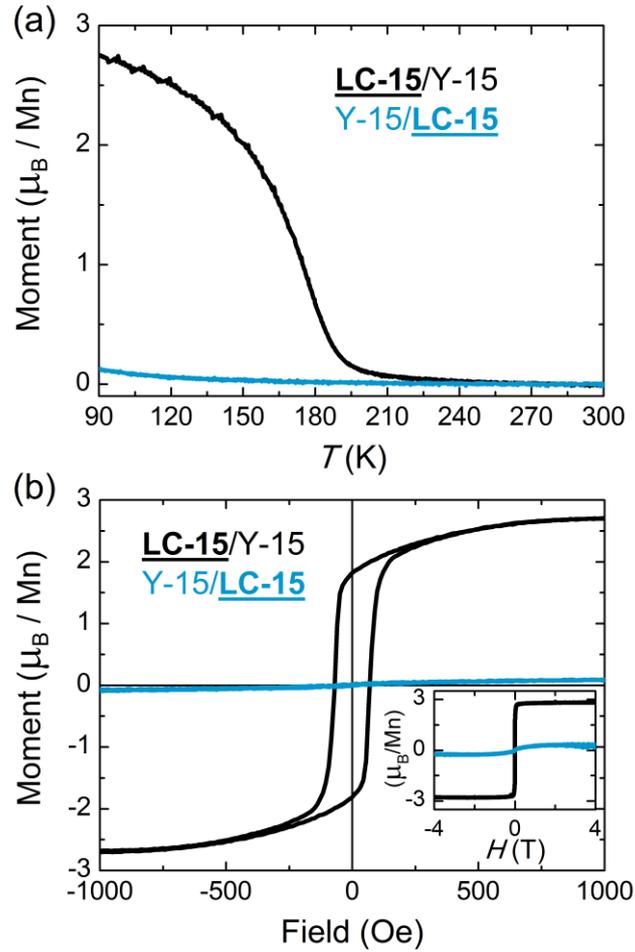

**Figure 5.** DC magnetisation of A-type grown bilayers LC-15/Y-15 and Y-15/LC-15 that differ only concerning the layer sequence. **(a)** *M-T* curves obtained while cooling in 1000 Oe parallel to the layers. A clear ferromagnetic transition is observed for LC-15/Y-15 but hardly for Y-15/LC-15, which has a strongly suppressed ferromagnetic moment. **(b)** *M-H* curves recorded at 100 K (in-plane field). Inset: same *M-H* curve for an expanded field range up to 4 T.

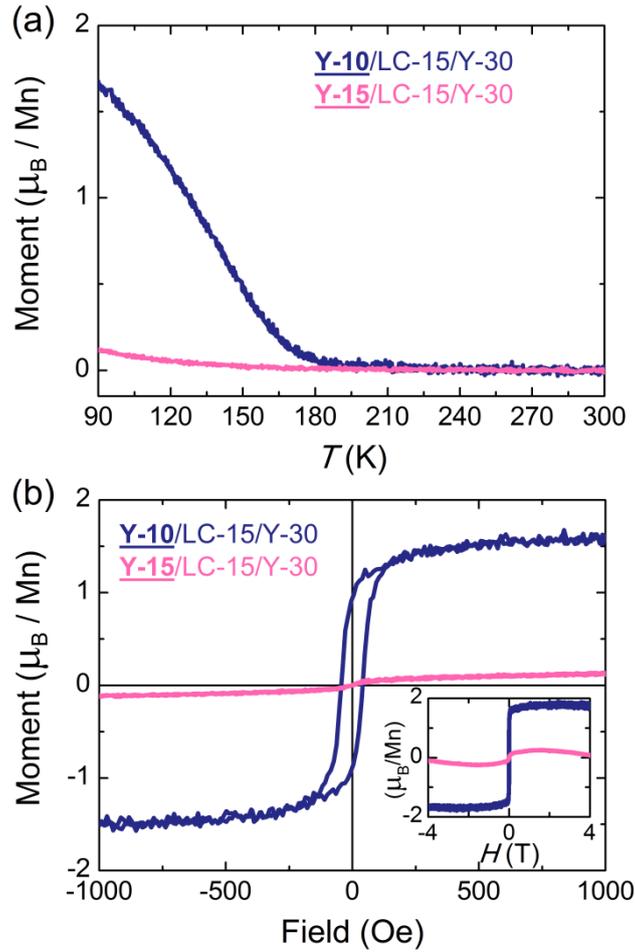

**Figure 6.** DC magnetisation of A-type-grown trilayers of Y-10/LC-15/Y-30 and Y-15/LC-15/Y-30 with different thickness of the bottom YBCO layer. **(a)** *M-T* curves obtained while cooling in 1000 Oe parallel to the layers. A clear ferromagnetic signal with $T^{Curie} \approx 180$ K is observed for the trilayer with the 10-nm thick bottom YBCO layer, but hardly for the one with the 15-nm thick bottom YBCO layer. **(b)** *M-H* curves recorded at 100 K (in-plane field). Inset: same *M-H* curve with expanded field range up to 4 T.

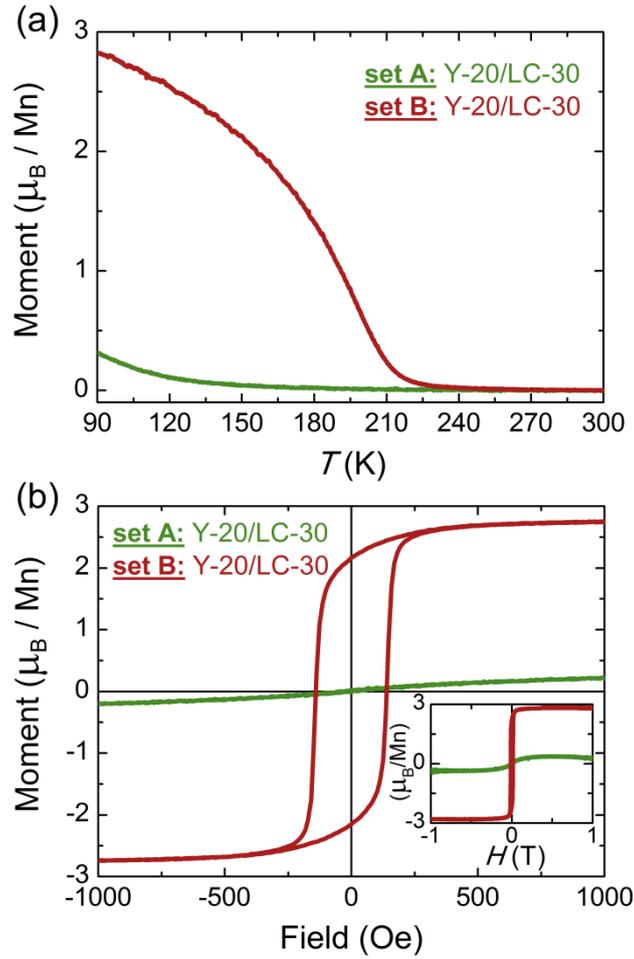

**Figure 7.** DC magnetisation for two Y-20/LC-30 bilayers grown under A-type (green lines) and B-type (red lines) conditions. **(a)** *M-T* curves obtained while cooling in 1000 Oe parallel to the layers. The A-type sample shows a strongly suppressed magnetism, while the B-type bilayer exhibits a sharp ferromagnetic transition around $T^{Curie} \approx 210$ K and a large moment of 2.8 $\mu_B$/Mn at 90 K. **(b)** *M-H* curve recorded at 100 K (in-plane field). Inset: same *M-H* curve with expanded field range up to 1 T.

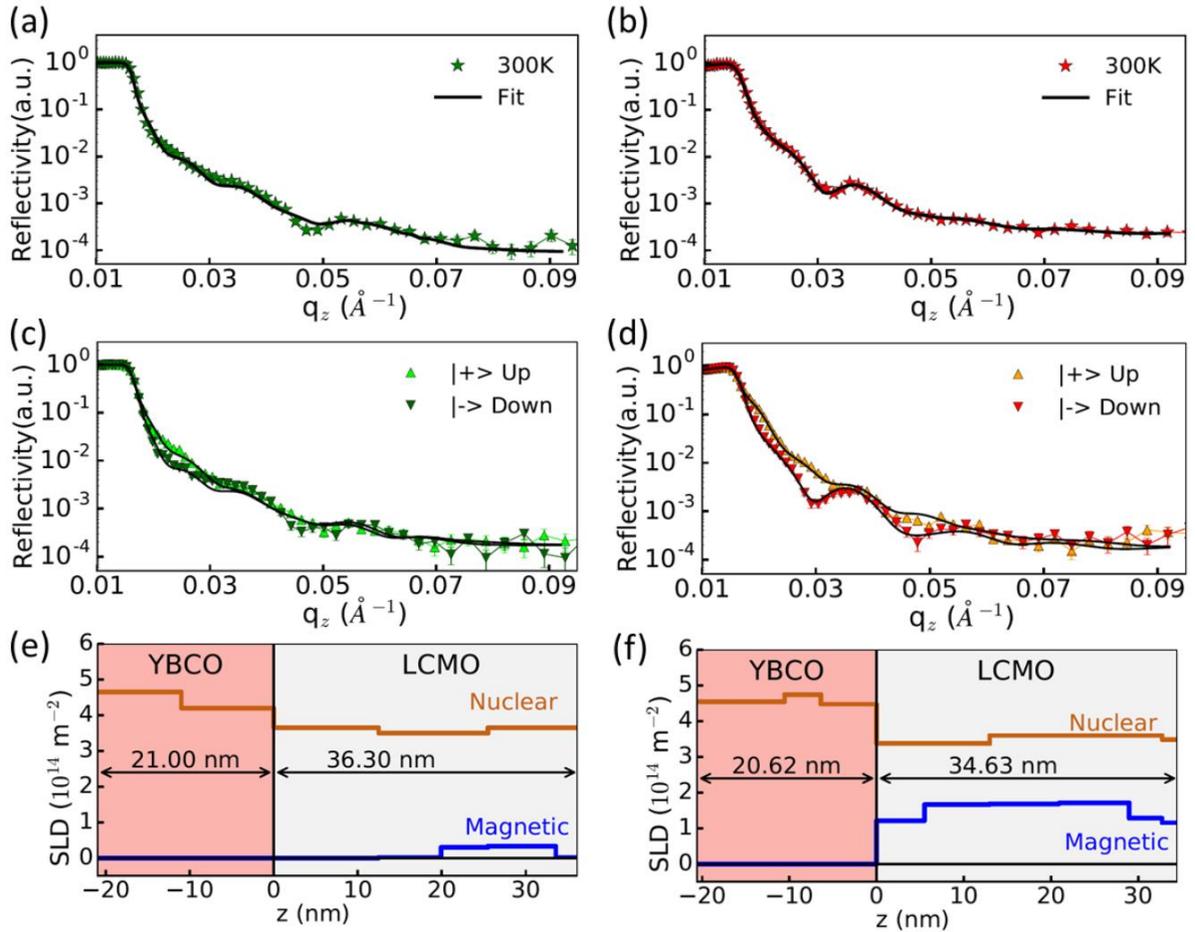

**Figure 8.** Polarized neutron reflectometry (PNR) curves of A- and B-type Y-20/LC-30 bilayers. **(a)** and **(b)** Reflectivity curves (symbols) and fits (lines) of the A- and B-type bilayers, respectively, obtained in the paramagnetic state at 300 K for which the spin-up and spin-down signals are identical. **(c)** and **(d)** Corresponding polarized reflectivity curves (symbols) and fits (lines) obtained at 90 K in the ferromagnetic state of LCMO. **(e)** and **(f)** Depth profiles of the nuclear (brown line) and magnetic (blue line) scattering length densities obtained from fitting the reflectivity curves of the A- and B-type bilayers, respectively.

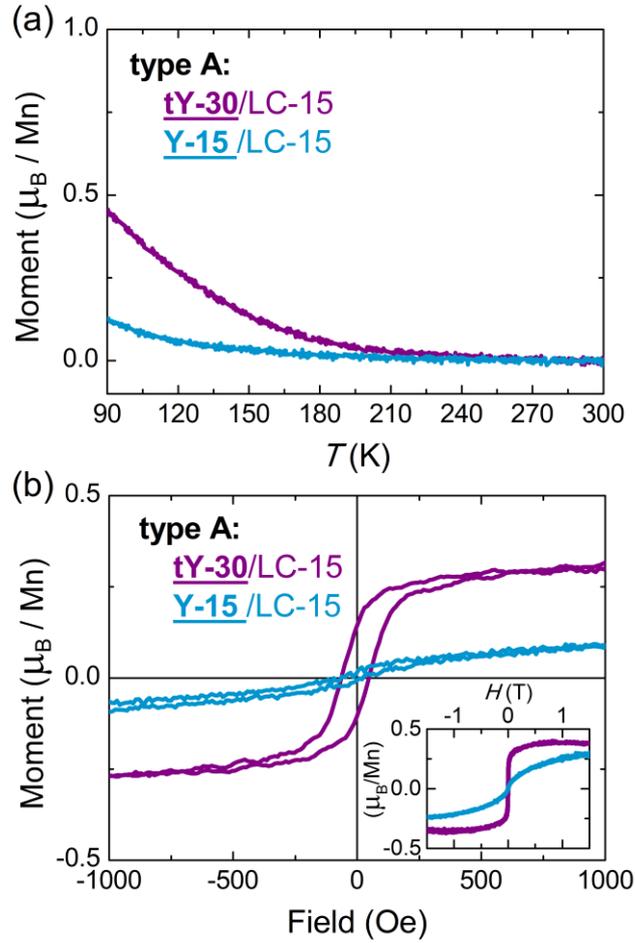

**Figure 9.** DC magnetisation data of the A-type bilayers tY-30/LC-15 and Y-15/LC-15. **(a)** *M-T* curves obtained while cooling in 1000 Oe parallel to the layers. The tY-30/LC-15 bilayer exhibits a larger ferromagnetic response than Y-15/LC-15 despite of having a two times thicker intermediate cuprate layer with $T^{Curie} \approx 180$ K. **(b)** *M-H* curves recorded at 100 K (in-plane field). Inset: same *M-H* curves with expanded field range up to 1.5 T.

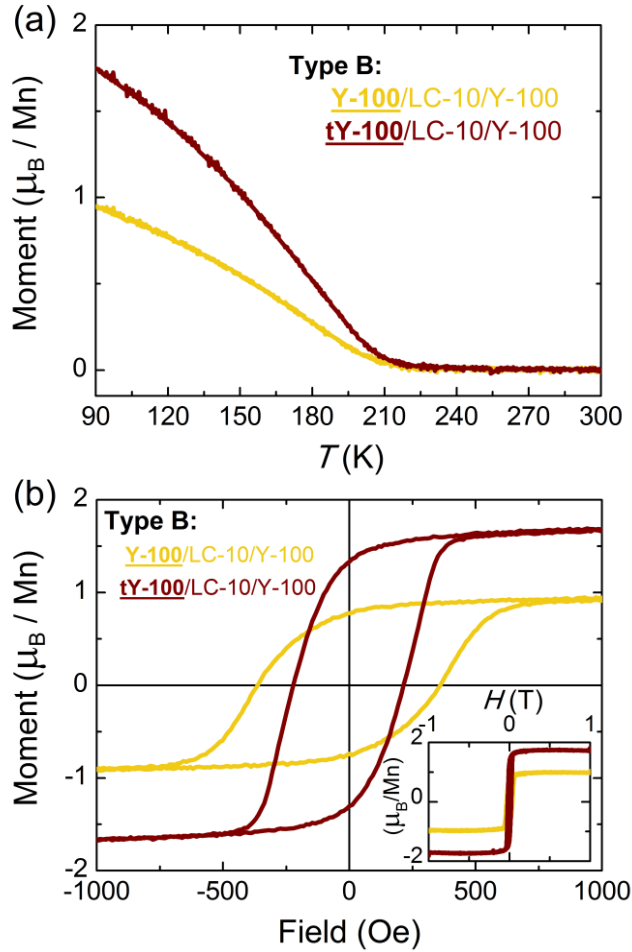

**Figure 10.** DC magnetisation data of B-type trilayers Y-100/LC-10/Y-100 and tY-100/LC-10/Y-100. **(a)** *M-T* curves obtained while cooling in 1000 Oe parallel to the layers. A transition to a ferromagnetic state is observed in both cases at $T^{Curie} \approx 200$ K, but a considerably larger magnetisation at 90 K is obtained for the tY-100/LC-10/Y-100 trilayer. **(b)** *M-H* curves recorded at 100 K (in-plane field). Inset: same *M-H* curves with expanded field range up to 1 T.

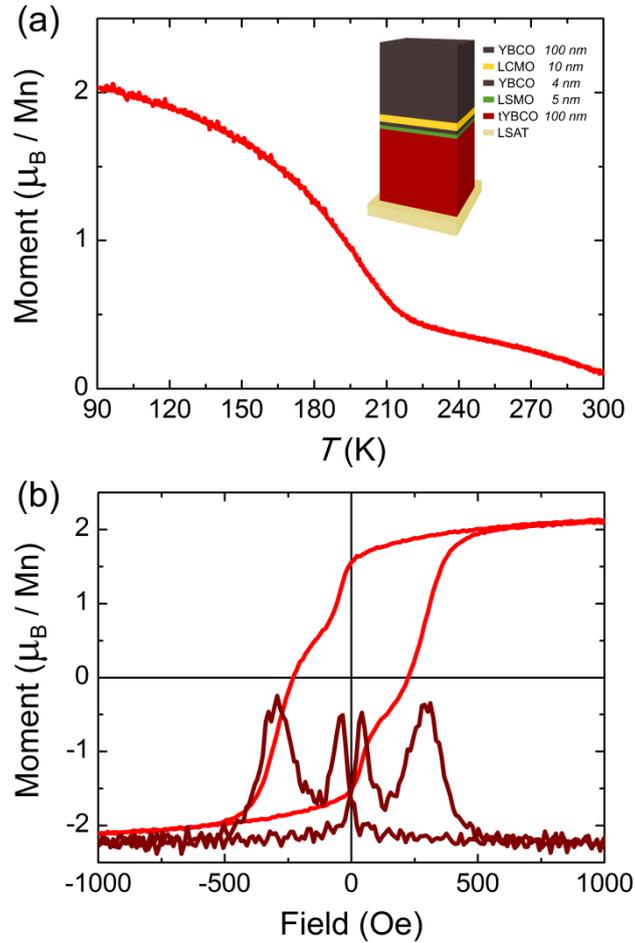

**Figure 11.** DC magnetisation data of a spin-valve-type multilayer with two very thin manganite layers. **(a)** *M-T* curve obtained while cooling in 1000 Oe parallel to the layers. A clear ferromagnetic transition due to the LCMO layer is observed at $T^{Curie} \approx 210$ K. The corresponding ferromagnetic transition of the LSMO layer occurs above 300 K. **(b)** *M-H* curves recorded at 90 K (in-plane field). The hysteresis-loop has a two-step transition that is most clearly seen in the corresponding derivative curve (darker line). The peaks of the latter mark the coercive fields of $H_{coer} \approx 40$ Oe and 300 Oe for LSMO and LCMO, respectively.